\documentclass[numbers]{article}

\usepackage[dblblindworkshop, final]{neurips_2025_libribrain_competition}

\workshoptitle{LibriBrain Competition 2025}

\usepackage[utf8]{inputenc} 
\usepackage[T1]{fontenc}    
\usepackage{hyperref}       
\usepackage{url}            
\usepackage{booktabs}       
\usepackage{amsfonts}       
\usepackage{nicefrac}       
\usepackage{microtype}      
\usepackage{xcolor}         
\usepackage{graphicx}
\usepackage{amsmath}
\usepackage{multirow}
\usepackage[table,xcdraw]{xcolor}

\title{Bypassing Direct Reconstruction: Speech Detection from MEG via Large-Scale Audio Retrieval}

\author{%
 Boda Xiao$^{a,c}$, Bo Wang$^{b,c}$, Heping Cheng$^{d*}$ \\
  $^{a}$Center for BioMed-X Research, Academy for Advanced Interdisciplinary Studies, \\
  Peking University, Beijing, China \\
  $^{b}$Speech and Hearing Research Center, School of Intelligence Science and Technology, \\
  Peking University, Beijing, China \\
  $^{c}$State Key Laboratory of General Artificial Intelligence, Beijing, China \\
  $^{d}$National Biomedical Imaging Center, State Key Laboratory of Membrane Biology, \\
  Institute of Molecular Medicine, Peking-Tsinghua Center for Life Sciences, \\
  College of Future Technology, Peking University, Beijing, China \\
  \texttt{chengp@pku.edu.cn}
}

\begin{document}

\maketitle

\begin{abstract}
Decoding speech from non-invasive brain signals is challenging. For the LibriBrain 2025 Speech Detection task, we propose a novel two-step framework that bypasses direct reconstruction. First, a contrastive learning model retrieves the matching speech segment for the given test MEG from a large-scale audio library (LibriVox). Second, a speech detection model generates the binary silence/speech sequence directly from this retrieved audio. With this approach, our team \textbf{Sherlock Holmes} achieved first place in the extended track (F1-score: 0.962), demonstrating that leveraging external audio databases is a highly effective strategy.
\end{abstract}

\section{Introduction}
Speech perception involves transforming auditory inputs into increasingly abstract language representations \cite{DeWitt_Rauschecker_2012, Heer_Huth_Griffiths_Gallant_Theunissen_2017, Poeppel_2014, Scott_Johnsrude_2003}. Accordingly, non-invasive magnetoencephalography (MEG) or electroencephalography (EEG) recordings during speech perception have been shown to capture hierarchical features of the speech \cite{Accou_Vanthornhout_Hamme_Francart_2023, Chan_Halgren_Marinkovic_Cash_2011, Defossez_Caucheteux_Rapin_Kabeli_King_2023, Di_Liberto_OSullivan_Lalor_2015, Ding_Melloni_Zhang_Tian_Poeppel_2016}. Numerous studies have successfully related M/EEG with speech. These efforts can be broadly categorized into two paradigms: regression tasks and match-mismatch tasks \cite{Puffay_Accou_Bollens_Monesi_Vanthornhout_Hamme_Francart_2023}. In regression tasks, neural networks are employed to reconstruct speech features directly from  M/EEG segments, such as envelopes, mel-spectrograms, and et al. \cite{Accou_Vanthornhout_Hamme_Francart_2023, Wang_Xu_Zhang_Xiao_Wu_Chen_2024, Xu_Wang_Yan_Zhu_Zhang_Wu_Chen_2024}. In match-mismatch tasks, neural networks learn to identify the target speech segment a subject is listening to from a predefined pool of candidates by maximizing the similarity between M/EEG segments and speech segments in a latent space \cite{Defossez_Caucheteux_Rapin_Kabeli_King_2023, Monesi_Accou_Montoya_Francart_Hamme_2020, Wang_Xu_Zhang_Zhu_Yan_Wu_Chen_2024a}. Collectively, these studies have made significant contributions toward developing non-invasive Brain-Computer Interfaces (BCIs) based on speech decoding.

In the LibriBrain Competition 2025 Speech Detection task, participants are required to train a model to distinguish between speech and silence based on brain activity measured by MEG \cite{Landau_Ozdogan_Elvers_Mantegna_Somaiya_Jayalath_Kurth_Kwon_Shillingford_Farquhar_2025}. In this setup, the label '0' corresponds to silence and '1' to speech. This task can be viewed as a regression problem, aiming to reconstruct a binary 0/1 sequence from MEG signals. However, reconstructing  dynamic speech features from M/EEG data is challenging due to the relatively low signal-to-noise ratio (SNR). For instance, the accuracy (measured by the Pearson correlation coefficient between decoded and target speech features) of decoding mel-spectrograms from EEG is typically below 0.2 \cite{Xu_Wang_Yan_Zhu_Zhang_Wu_Chen_2024, Li_Fang_Zhang_Chen_Gao_2024, Sakthi_Tewfik_Chandrasekaran_2019}. Our own experiments indicate that even with the superior signal quality of MEG, the accuracy for decoding mel-spectrograms remains around 0.4. This level of accuracy is currently insufficient to support the synthesis of intelligible speech. In contrast, match-mismatch tasks benefit from the constraint provided by the candidate set. Previous research has shown that models can identify the target speech segment from a pool of over 1000 candidates based on 3-second MEG recordings, achieving an average accuracy of 41\%, which highlights a promising avenue for decoding speech from non-invasive brain activity \cite{Defossez_Caucheteux_Rapin_Kabeli_King_2023}.

\begin{figure}[htbp]
    \centering
    \includegraphics[width=0.6\textwidth]{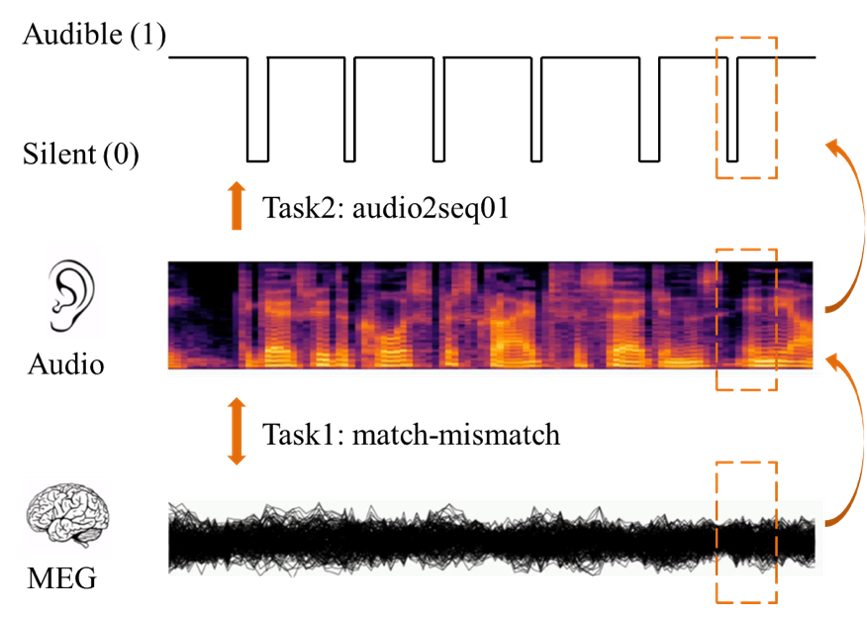}
    \caption{Overall framework of our approach.}
    \label{fig:my_label}
\end{figure}

Inspired by these findings, we proposed a two-step decoding approach for the extended track of the Speech Detection task, as illustrated in Figure ~\ref{fig:my_label}. In the first step, we employed a match-mismatch task to identify the speech segment corresponding to the test MEG, from a large-scale dataset LibriVox. In the second step, we performed the speech detection task on the matched speech segment. Our approach ultimately achieved first place on the track.

\section{Methods}

\subsection{Step1: MEG-Speech Match-mismatch}
\begin{figure}[htbp]
    \centering
    \includegraphics[width=0.8\textwidth]{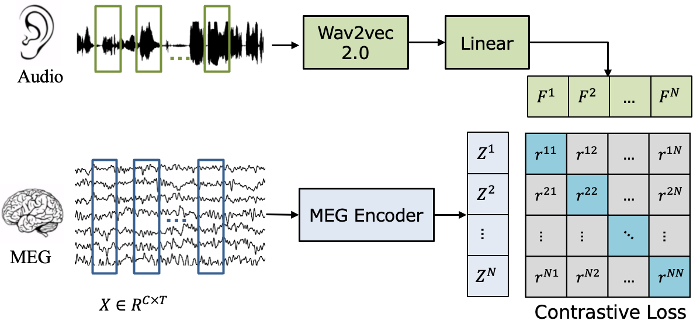}
    \caption{The contrastive learning framework for the match-mismatch task.}
    \label{fig:framework}
\end{figure}

The objective of this step is to align MEG recordings with speech segments. We adopted a contrastive learning framework, as depicted in Figure ~\ref{fig:framework}. For a segment of MEG data $X\in \mathbb{R}^{C \times T}$, where $C$ represents the number of MEG channels and $T$ represents the time samples. A CNN-based MEG encoder was utilized for extracting neural features $Z\in \mathbb{R}^{H \times T}$ . Meanwhile, a pretrained Wav2vec 2.0 model \footnote{We used wav2vec2-base-960h from \url{https://huggingface.co/facebook/wav2vec2-base-960h}} was used to obtain the speech representation (extracted from the outputs of its ninth hidden layer). This representation is subsequently projected via a linear layer to obtain features $F\in \mathbb{R}^{H \times T}$. Given a batch of $N$ samples, let $\mathcal{Z} = \{ Z^1, Z^2, \ldots, Z^N \}$ denote the MEG features and $\mathcal{F} = \{ F^1, F^2, \ldots, F^N \}$represent the speech representations. The InfoNCE (Information Noise-Contrastive Estimation) loss is employed \cite{Oord_Li_Vinyals_2018}, aiming to maximize the similarity between matched pairs $(Z^i,F^i)$ while minimizing the similarity between mismatched pairs $(Z^i,F^j)$ for $j\neq i$. The loss is formulated as:
\begin{equation}
\mathcal{L}_\text{InfoNCE} = -\frac{1}{N} \sum_{i=1}^N \log \frac{\exp(\text{sim}(Z^i, F^i)/\tau)}{\sum_{j=1}^N \exp(\text{sim}(Z^i, F^j)/\tau)}
\label{eq:infonce}
\end{equation}
where $\text{sim}(\cdot, \cdot)$ denotes the similarity measure, and $\tau$ is a temperature parameter that modulates the sharpness of the distribution. 
The $\text{sim}(Z^i, F^i)$ is calculated as:
\begin{equation}
\text{sim}(Z^i, F^i) = \frac{1}{H} \sum_{k=1}^H \text{corr}(z_k^i, f_k^i)
\label{eq:sim}
\end{equation}
where $\text{corr}(\cdot, \cdot)$ is the Pearson correlation between two vectors.

\subsection{Step2: Speech detection}
In this step, we train a speech detection model. This model takes the mel-spectrogram of a speech segment as input and outputs a binary sequence (0 for silence, 1 for speech). The model is implemented using a deep CNN. The network parameters are optimized using the negative Pearson correlation coefficient as the loss function.

\section{Experiment}
\subsection{Data Preparation}
\label{sec:data_preparation}

Our framework requires MEG data and its temporally aligned speech signals for training. Following the speech source URLs provided in the organizers' paper \cite{Ozdogan_Landau_Elvers_Jayalath_Somaiya_Mantegna_Woolrich_Jones_2025}, we downloaded all corresponding audiobooks from LibriVox, which we refer to as Libriaudio. The actual audio stimuli presented to subjects during the MEG recording sessions are denoted as MEGaudio. Analysis of the dataset's event.tsv file, specifically by comparing the 'timemeg' and 'timechapter' columns, revealed that MEGaudio was generated by inserting silent segments into the original Libriaudio. For instance, in the session 'sub-0\_ses-1\_task-Sherlock1\_run-1\_proc-bads+headpos+sss+notch+bp+ds\_meg.h5', the corresponding MEGaudio contained 171 additional silent segments compared to the original Libriaudio. The duration distribution of these extra silent segments is shown in Figure~\ref{fig:audio_synthesis}a (median duration $\approx$ 0.03 s). Cumulatively, these segments added approximately 5 seconds of silence to the MEGaudio. As illustrated in Figure~\ref{fig:audio_synthesis}b, we used the timing information from the event.tsv file to synthesize the MEGaudio by inserting silent segments of corresponding durations into the Libriaudio at the specified timestamps. This synthesized MEGaudio was subsequently used for training both the MEG-Speech match-mismatch model and the Speech Detection model.

\subsection{Model Training}
Data from session 9 and 10 of the Sherlock 1 were used as the local validation set, while data from session 11 and 12 served as the local test set. All remaining data constituted the training set.

For the MEG-speech match-mismatch model, the ConvConcatNet architecture \cite{Xu_Wang_Yan_Zhu_Zhang_Wu_Chen_2024} was employed as the MEG encoder. The dimensionality of the latent space was set to 8 to reduce computational cost during the subsequent testing phase. Following previous work, the MEG data and corresponding speech for each session were segmented into non-overlapping 3-second windows. The model was trained using the Adam optimizer \cite{Kingma_Ba_2015} with a learning rate of $1 \times 10^{-3}$ and a batch size of 256. The temperature parameter $\tau$ in the InfoNCE loss was set to 0.015. Training was stopped if the Top-10 accuracy on the validation set failed to improve for 5 consecutive epochs.

For the speech detection model, the ConvConcatNet was also used. The model input was the mel-spectrogram of the MEGaudio, and the target labels (binary 0/1 sequences) were derived from the event.tsv file. Segment length was set to 30 seconds to ensure that each segment contained both speech and silence periods. The model was trained using the Adam optimizer with a learning rate of $1 \times 10^{-3}$ and a batch size of 64. The negative Pearson correlation coefficient was used as the loss function. Training was stopped if the validation loss did not decrease for 5 consecutive epochs. The optimal binarization threshold for distinguishing speech from silence was determined via a grid search on the validation set.

All models were trained on an HPC node equipped with 8 A800 GPUs.
\subsection{Model Testing}
\subsubsection{Retrieving Matching Speech from LibriVox}
During testing, our goal was to retrieve the speech segment from the LibriVox corpus that matched the given test MEG. We downloaded a large-scale subset of LibriVox, comprising approximately 60\% of its total data ($\sim$10,000 audiobooks). Each audiobook was split into non-overlapping 5-second segments. As an example, the chapter ``A Continuation of the Reminiscences of John Watson MD'' from ``A Study In Scarlet (Version 6)'' (hereafter \textit{studyinscarlet13}), with a duration of 27 minutes and 31 seconds, was split into 330 segments.

The holdout test MEG data had a total duration of 2243 seconds. It was segmented using a 5-second sliding window with a 0.1-second stride, resulting in 22,380 MEG segments. The matching process is illustrated in Figure~\ref{fig:mmis_pattern}a. For each of the 330 speech segments from \textit{studyinscarlet13}, we identified the most similar MEG segment from the pool of 22,380 test segments, recording its index. This produced a sequence of 330 indices, which we term the Matched MEG ID Sequence (MMIS). For the vast majority of LibriVox audiobooks, which do not correspond to the holdout MEG data, the MMIS shows no discernible pattern. However, for the matching audio, a large subset of the MMIS should form a monotonically increasing sequence, reflecting the temporal order of the MEG data (Figure~\ref{fig:mmis_pattern}b).

To identify this subset, we computed the Longest Ascending Subsequence (LAS) of the MMIS. Experimental results indicated that among the $\sim$10,000 downloaded audiobooks, only \textit{studyinscarlet13} yielded an LAS length exceeding a manually set threshold of 20. This audio was identified as matching the final portion of the holdout test MEG, starting from the 1398-s mark. No matching audio was found for the MEG data preceding 1398 s.

\subsubsection{Generating the Binary Sequence from Speech}
Based on the analysis in Section~\ref{sec:data_preparation}, which indicated that most extra silent segments were inserted between sentences, we segmented the \textit{studyinscarlet13} audio into 241 sentences according to its text transcript. Using the trained MEG-Speech match-mismatch model, the first 126 sentences were confirmed to match the test MEG data. Silent segments of corresponding durations were inserted between these sentences so that the total duration matched that of the MEG data after 1398 s. Finally, the trained speech detection model was applied to generate the binary 0/1 sequence, which served as the decoded output for the MEG signal after 1398 s.

For the initial 1398 s of the test MEG, no audio file from our LibriVox subset produced an MMIS with an LAS length greater than 20. We hypothesize that the corresponding audio might reside in the remaining 40\% of the LibriVox corpus we did not download, or originate from another source outside LibriVox. An interesting observation was that segments from audiobook ``The Darkest Hour'' appeared frequently in matches for the preceding 1398 s, but not in sequential order. For this initial portion, we employed a simple regression approach, similar to the method used by Team SHINE (as discussed in the provided Discord thread), utilizing a basic CNN+LSTM network to reconstruct the binary sequence directly from the MEG signals.

Submitting the prediction result comprising of the above two parts to the extended track, we obtained an F1-score of 0.962, securing first place on the leaderboard.

\section{Conclusion}
We presented a two-step framework for speech detection from MEG signals. By reframing the problem as an audio retrieval task followed by speech analysis, we circumvented the limitations of direct feature regression from noisy neural data. Our method successfully identified the target audio from a vast pool of candidates and generated accurate binary sequences, winning the LibriBrain competition. This work validates the potential of using match-mismatch tasks to advance non-invasive BCIs.

\begin{ack}
This work was supported by the STI 2030--Major Projects (No.2021ZD0201500), the High-performance Computing Platform of Peking University, and the Biomedical Computing Platform of National Biomedical Imaging Center of Peking University. We also gratefully acknowledge the guidance and valuable discussions provided by Prof. Jing Chen.
\end{ack}

\bibliographystyle{IEEEtran}
\bibliography{shylock}


\appendix
\setcounter{figure}{0}
\renewcommand{\thefigure}{A.\arabic{figure}}
\section{Supplementary Material}

\begin{figure}[htbp]
    \centering
    \includegraphics[width=0.8\textwidth]{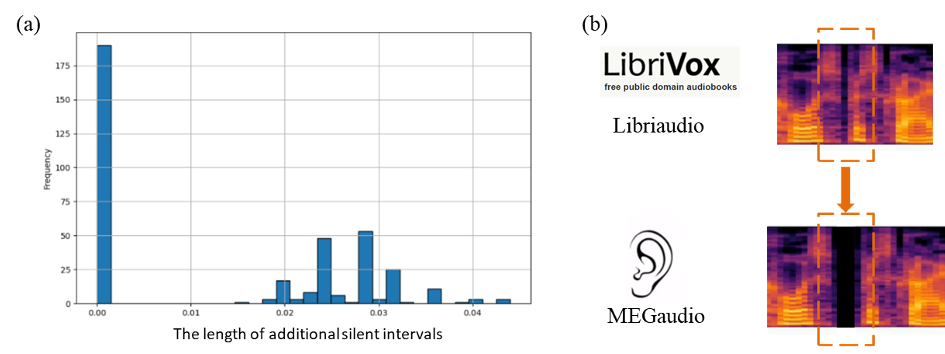}
    \caption{(a) Duration distribution of the extra silent segments for an example session. (b) Synthesizing MEGaudio by inserting silent segments into the original Libriaudio.}
    \label{fig:audio_synthesis}
\end{figure}

\begin{figure}[htbp]
    \centering
    \includegraphics[width=0.8\textwidth]{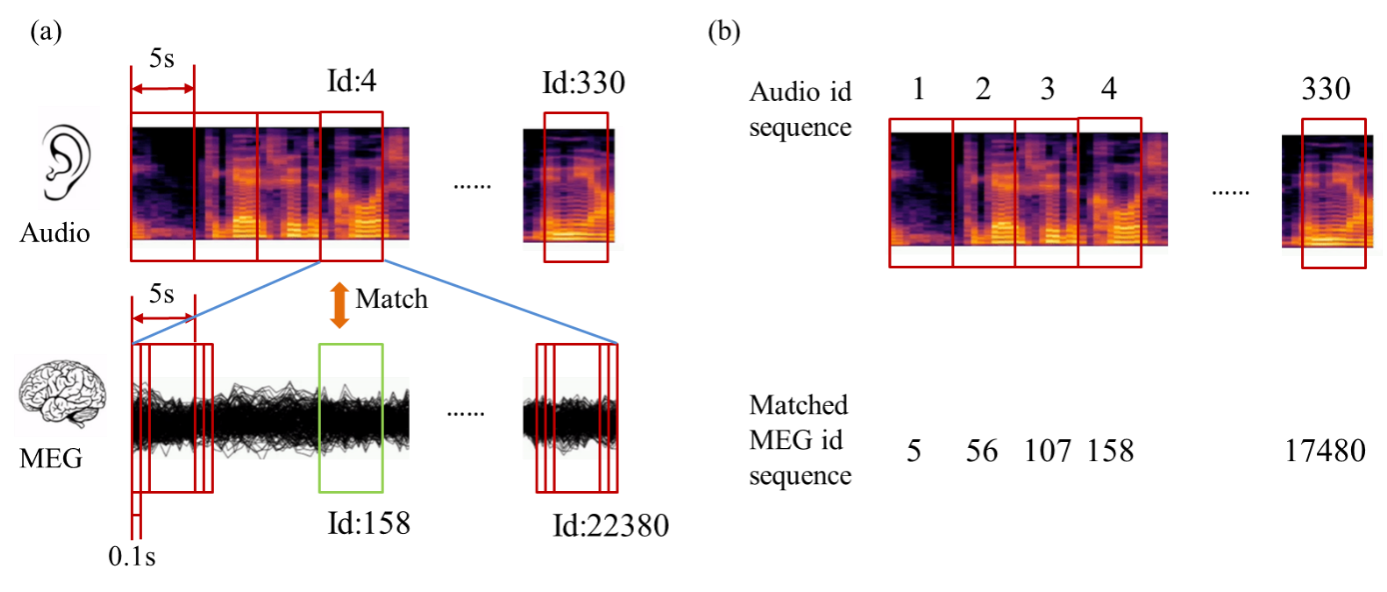}
    \caption{(a) During testing, speech segments from the LibriVox corpus are matched against segments from the holdout MEG data. (b) The ideal MMIS for the matching audio is a monotonically increasing sequence.}
    \label{fig:mmis_pattern}
\end{figure}


\end{document}